\documentstyle[pre,aps,epsfig]{revtex}
\begin{document}
\title{Dynamics of inelastically colliding rough spheres: Relaxation
        of translational and rotational energy}

\author{Martin Huthmann and
        Annette   Zippelius} 
\address{Institut f\"ur Theoretische Physik, 
Georg--August--Universit\"at, 37073 G\"ottingen, Germany}
\date{\today}
\maketitle

\begin{abstract}
We study the exchange of kinetic energy between translational and
rotational degrees of freedom for inelastic collisions of rough
spheres. Even if equipartition holds in the initial state it is
immediately destroyed by collisions. The simplest generalisation of 
the homogeneous cooling state allows for two temperatures, characterizing
translational and rotational degrees of freedom separately. For times
larger than a crossover frequency, which is determined by the Enskog
frequency and the initial temperature, both energies decay
algebraically like $t^{-2}$ with a fixed ratio of amplitudes,
different from one. 
\end{abstract}
\vspace*{-20pt}

 Kinetic theory of inelastically colliding particles has become a
 subject of growing research activity, motivated partly by renewed
 interest in granular materials. A Boltzmann equation and the Enskog
 variant of it have been formulated for inelastically colliding
 particles with normal and tangential restitution
 \cite{Jenkins,Savage,Richman,Lun,Goldshtein,Cercignani}. Whereas the
 derivation of the kinetic equation is relatively straightforward, the
 methods of solution which were developped for elastic collisions
 cannot be taken over to the inelastic case because there is no simple
 stationary or local equilibrium distribution, around which one could
 expand the nonlinear kinetic equation. One simple distribution is the
 homogeneous cooling state (HCS)\cite{McNamara,Esipov}, which depends
 on time only implicitly via the average kinetic energy $T(t)$. The
 latter is predicted to decay like $t^{-2}$ for large times. The
 homogeneity assumption is certainly violated when clustering occurs
 and, in particular, if inelastic collapse happens. Nevertheless we have
 found recently\cite{Aspelmeier} in a model of inelastically colliding
 rods that the kinetic energy follows {\it on average} a $t^{-2}$
 behaviour, even if clustering occurs. This suggests that the above
 scaling law may be a useful approximate description, even when the
 assumptions of HCS break down.

 In this note we show that rotational and translational energy relax
 with different rates in general. Once friction is included, HCS with a
 single time dependent temperature is no longer consistent with the 
 time evolution of translational and rotational energy
 separately. Instead one has to introduce two temperatures which
 characterize translational and rotational degress of freedom
 separately. Both are found to fall off like $t^{-2}$ with the ratio
 approaching a constant value, which is determined by the coefficients
 of normal and tangential restitution.

 We briefly recall the collision dynamics of hard
 spheres with normal and tangential restitution. These results can for
 example be found in Cerginani \cite{Cercignani}.
  We consider two spheres of equal 
 diameter $a$, mass $M$ and moment of inertia $I$. The unit-vector 
 from the center of the first sphere to the center of the second is
 denoted by $\hat{\mbox{\boldmath$n$}}$ and velocities and angular velocities
 before collision by
 $\mbox{\boldmath$v$}_1$, $\mbox{\boldmath$v$}_2$, $\mbox{\boldmath$\omega$}_1$ and $\mbox{\boldmath$\omega$}_2$.
 The relative velocity of the contactpoint before collision is given by
 $  \mbox{\boldmath$V$}=\mbox{\boldmath$v$}_2 +\frac{a}{2}\hat{\mbox{\boldmath$n$}}\times \mbox{\boldmath$\omega$}_2-
   \mbox{\boldmath$v$}_1 +\frac{a}{2}\hat{\mbox{\boldmath$n$}}\times \mbox{\boldmath$\omega$}_1$ .
 Normal and tangential restitution determine the relative velocity
 after collision according to
 \begin{eqnarray}\label{rauku4}
   \hat{\mbox{\boldmath$n$}}\cdot\mbox{\boldmath$V$}^{'}
&=& -\epsilon(\hat{\mbox{\boldmath$n$}}\cdot\mbox{\boldmath$V$})\
   {\rm with}
   \quad \epsilon \in [0,1] ,
\\ \label{rauku5}
   \hat{\mbox{\boldmath$n$}}\times\mbox{\boldmath$V$}^{'}
&=& -\beta(\hat{\mbox{\boldmath$n$}}\times\mbox{\boldmath$V$})\ 
   {\rm with}
   \quad \beta \in [-1,1] .
 \end{eqnarray}
 Using the property of conserved linear and angular momenta one obtains
 for the velocities after collision 
 \begin{eqnarray} \nonumber
   \mbox{\boldmath$v$}_1^{'}
&=&
  \mbox{\boldmath$v$}_1-\eta_t\mbox{\boldmath$v$}_{12} -
   (\eta_n-\eta_t) 
   (\hat{\mbox{\boldmath$n$}}\cdot\mbox{\boldmath$v$}_{12})
   \hat{\mbox{\boldmath$n$}}+\eta_t\frac{a}{2} 
   \hat{\mbox{\boldmath$n$}}\times(\mbox{\boldmath$\omega$}_1+\mbox{\boldmath$\omega$}_2),\\ \nonumber
   \mbox{\boldmath$v$}_2^{'}
&=&
   \mbox{\boldmath$v$}_2+\eta_t\mbox{\boldmath$v$}_{12}+
   (\eta_n-\eta_t)
   (\hat{\mbox{\boldmath$n$}}\cdot\mbox{\boldmath$v$}_{12})
   \hat{\mbox{\boldmath$n$}}-\eta_t\frac{a}{2}
   \hat{\mbox{\boldmath$n$}}\times(\mbox{\boldmath$\omega$}_1+\mbox{\boldmath$\omega$}_2),\\ \nonumber
   \mbox{\boldmath$\omega$}_1^{'} 
&=& 
   \mbox{\boldmath$\omega$}_1-\frac{2}{a k} \eta_t  
   \hat{\mbox{\boldmath$n$}}\times\mbox{\boldmath$v$}_{12}
   +\frac{\eta_t}{k} \hat{\mbox{\boldmath$n$}}\times
   (\hat{\mbox{\boldmath$n$}}\times(\mbox{\boldmath$\omega$}_1+\mbox{\boldmath$\omega$}_2)),\\ \label{freddy}
   \mbox{\boldmath$\omega$}_2^{'} 
&=&
   \mbox{\boldmath$\omega$}_2-\frac{2}{a k} \eta_t  
   \hat{\mbox{\boldmath$n$}}\times\mbox{\boldmath$v$}_{12}
   +\frac{\eta_t}{k} \hat{\mbox{\boldmath$n$}}\times
   (\hat{\mbox{\boldmath$n$}}\times(\mbox{\boldmath$\omega$}_1+\mbox{\boldmath$\omega$}_2)).
 \end{eqnarray}
 with $\mbox{\boldmath$v$}_{12}=\mbox{\boldmath$v$}_1-\mbox{\boldmath$v$}_2 $ and parameters $ k:=
 \frac{4I}{M a^2}$ (k=0.4 for homogeneous spheres), $\eta_n:=\frac{1+\epsilon}{2}$, and
 $\eta_t:=\frac{k(1+\beta)}{2(1+k)}$.

 We consider a system of N classical particles, confined to a
 3-dimensional volume $V$ and interacting via a hard 
 core potential. Each particle is
 characterized by its position $\mbox{\boldmath$r$}_i(t)$, its linear momentum
 $\mbox{\boldmath$p$}_i(t)=M \mbox{\boldmath$v$}_i(t)$ and angular
 velocity $\mbox{\boldmath$\omega$}_i(t)$. 
 The time development of a dynamical variable
 $A=A(\{\mbox{\boldmath$r$}_i(t),\mbox{\boldmath$p$}_i(t),\mbox{\boldmath$\omega$}_i(t)\})$
 is determined by a Pseudo Liouville--operator ${\cal L}_{+}$ 
 \begin{equation}
   A(\{\mbox{\boldmath$r$}_i,\mbox{\boldmath$p$}_i,\mbox{\boldmath$\omega$}_i\},t) =\exp(i{\cal
   L}_{+}t)A(\{\mbox{\boldmath$r$}_i,\mbox{\boldmath$p$}_i,\mbox{\boldmath$\omega$}_i\},0)\quad {\rm for}\quad t>0. 
 \end{equation}
 Such a Pseudo Liouville--operator was first introduced for hard
 (perfectly smooth) spheres by Ernst et al.\cite{Ernst} and
 subsequently applied to the
 calculation of transport coefficients of a hard sphere fluid 
 (see e.g. \cite{Yip} and references therein). Noije and Ernst\cite{Noije} have generalised the
 formalism to inelastic collisions with normal restitution. Here we
 extend these results to rough spheres with normal and tangential restitution.

 The Pseudo Liouville--operators ${\cal L}_{+}$ consists of two parts
 ${\cal L}_{+}={\cal L}_0
 +{\cal L}^{'}_{+}$. The first one, ${\cal L}_0$, describes free streaming 
 of particles
 \begin{equation}
   {\cal L}_0 = -\frac{i}{M} \sum_n 
   \mbox{\boldmath$p$}_n \cdot {\nabla}_{\mbox{\boldmath$\scriptstyle r$}_n}
 \end{equation}
 and the second one, ${\cal L}^{'}_{+}=\frac{1}{2}\sum_{n\ne m}T_{+}(nm)$, describes hard-core 
 collisions of two particles by
 \begin{equation} 
 \label{stossit}
          T_{+}(nm) = 
     \frac{i}{M}(\mbox{\boldmath$p$}_{nm}\cdot\hat{\mbox{\boldmath$r$}}_{nm})
    \Theta(-\mbox{\boldmath$p$}_{nm}\cdot\hat{\mbox{\boldmath$r$}}_{nm} )
   \delta(|\mbox{\boldmath$r$}_{nm}|-a)(b_{nm}^{+}-1). 
  \end{equation}
 The operator $b_{nm}^+$ replaces the linear and angular
 momenta of two particles $n$ and $m$ before collision 
 by the corresponding ones after collision.
 $\Theta(x)$ is the
 Heaviside step--function and we have introduced the notation 
 $\mbox{\boldmath$p$}_{nm}=\mbox{\boldmath$p$}_n-\mbox{\boldmath$p$}_m$,
 $\mbox{\boldmath$r$}_{nm}=\mbox{\boldmath$r$}_n-\mbox{\boldmath$r$}_m$ and $\hat{\mbox{\boldmath$r$}} =
 \mbox{\boldmath$r$}/|\mbox{\boldmath$r$}|$.

 Eq. (\ref{stossit}) has a simple interpretation. The factor
 $(\mbox{\boldmath$p$}_{nm}\cdot\hat{\mbox{\boldmath$r$}}_{nm})/M$ gives the flux of incoming
 particles. The $\Theta$- and $\delta$-functions specify the conditions
 for a collision to take place. A collision between particles $n$ and
 $m$ happens only if the two particles are approaching each other, which
is taken into account by
 $\Theta(-\mbox{\boldmath$p$}_{nm}\cdot\hat{\mbox{\boldmath$r$}}_{nm} )$. At the instant of a
 collision the distance between the two particles has to vanish, as
 expressed by $\delta(|\mbox{\boldmath$r$}_{nm}|-a)$. Finally $(b_{nm}-1)$ 
 generates the change of linear and angular momenta~\cite{fuss}.

 The ensemble average of a dynamical variable is defined by
 \begin{equation}
 <A>_t=\int d\Gamma \rho (0) A(t)=\int d\Gamma \rho (t) A(0)=
 \int \prod_i(d\mbox{\boldmath$r$}_i d\mbox{\boldmath$p$}_i d\mbox{\boldmath$\omega$}_i)
 \prod_{i<j}\Theta(|\mbox{\boldmath$r$}_{ij}|-a) \rho (t) A(0).
 \end{equation}
 Here $\rho(t)=\exp {(-i{\cal L}_+^{\dagger}t)}\,\rho(0)$ is the $N$-particle
 distribution function, whose time evolution is governed by the
 adjoint ${\cal L}_+^{\dagger}$ of the time evolution 
 operator ${\cal L}_+$. The
 quantities of interest are the translational and rotational energies per
 particle
 \begin{eqnarray}
 E_{\rm tr} & = & \frac{1}{N}\sum_i\frac{M}{2}\mbox{\boldmath$v$}_i^2\\
 E_{\rm rot} & = & \frac{1}{N}\sum_i\frac{I}{2}\mbox{\boldmath$\omega$}_i^2
 \end{eqnarray}
 as well as the total kinetic energy $E=E_{\rm tr}+E_{\rm rot}$.

 As a first step we compute the initial decay rates
 \begin{eqnarray}
 \frac{d}{dt}<E_{\rm tr}>_{t=0} & = & <i{\cal L}_{+}E_{\rm tr}>_{t=0}=\nu_{\rm tr}\\
 \frac{d}{dt}<E_{\rm rot}>_{t=0} & = & <i{\cal L}_{+}E_{\rm rot}>_{t=0}=\nu_{\rm rot}
 \end{eqnarray}
 assuming that the system has been prepared in a thermal equilibrium
 state $\rho(0) \propto \exp (-E/T)$ which is characterized by the 
 temperature $3T/2=<E>_{t=0}$. The change of translational and rotational
 energy are given by
 \begin{eqnarray} 
 \label{frequency}
 \nu_{\rm tr} & = & -\frac{2\pi a^2 T^{3/2}n_0 g(a)}{(\pi M)^{1/2}}
 \left( (1-\epsilon^2)+\frac{k}{1+k}(1-\beta^2)\right)\\
 \nu_{\rm rot} & = & -\frac{2\pi a^2 T^{3/2}n_0 g(a)}{(\pi M)^{1/2}}
 \left( \frac{1}{1+k}(1-\beta^2)\right).
 \end{eqnarray}
 Here $g(a)$ denotes the pair correlation at contact and
 $n_0=N/V$. Rotational energy is conserved in two cases, for either
 perfectly smooth spheres ($\beta=-1$) or perfectly
 rough spheres ($\beta=+1$). Translational energy is only conserved if
 in addition $\epsilon=1$. For all other
 values of the parameters $\epsilon$ and $\beta$ the translational and
 rotational energy decrease linearly with time but  with  
 {\it different} rates. This implies that after a small time interval
 $\Delta t$  equipartition among rotational and translational degrees
 of freedom no longer holds
 \begin{eqnarray}
 \label{inf.increase}
 <E_{\rm tr}>_{\Delta t} & = & <E_{\rm tr}>_0-\nu_{\rm tr}\Delta t\nonumber\\
 <E_{\rm rot}>_{\Delta t} & = & <E_{\rm rot}>_0-\nu_{\rm rot}\Delta t.
 \end{eqnarray}
 Hence for collisions with friction ($\beta\neq -1$) the homogeneous
 cooling state is not consistent with the time evolution of
 translational and rotational energy separately. To generalise
 the concept of a homogeneous cooling state to collisions with friction
 we introduce two temperatures $T_{\rm tr}(t)=2/3<E_{\rm tr}>_t$ and
$T_{\rm rot}(t)=2/3<E_{\rm rot}>_t$. We keep the assumption of spatial
homogeneity and assume that both linear and angular momenta are normally
distributed with in general different time dependent widths or
temperatures
\begin{equation}
\rho(t)\propto \exp{-\frac{1}{2}\sum_{i=1}^N (M \mbox{\boldmath$v$}_i^2/T_{\rm tr}(t)+
I\mbox{\boldmath$\omega$}_i^2/T_{\rm rot}(t))}.
\end{equation}
The above distribution allows for a calculation of the dacay rates of
 $T_{\rm tr}(t)$ and $T_{\rm rot}(t)$ for arbitrary times, resulting
in two coupled differential equations:
\begin{eqnarray}\label{coupletr}
\frac{d}{dt}T_{\rm tr} & = & -(1-\epsilon^2) \frac{\gamma}{4}T_{\rm tr}^{3/2}
-\gamma\eta_t\left(1-\eta_t\right)T_{\rm tr}^{3/2}
+\frac{\gamma}{k}\eta_t^2T_{\rm tr}^{1/2}T_{\rm rot}\\
\frac{d}{dt}T_{\rm rot} & = & \frac{\gamma}{k}
\label{couplerot}
\eta_t^2T_{\rm tr}^{3/2}
-\gamma\frac{\eta_t}{k}\left(1-\frac{\eta_t}{k}\right)
T_{\rm rot}T_{\rm tr}^{1/2}
\end{eqnarray}
with $\gamma=(16/3) a^2 n_0 g(a)\sqrt{\pi/M}$.

\begin{figure}[hbt]
\center  \epsfig{file=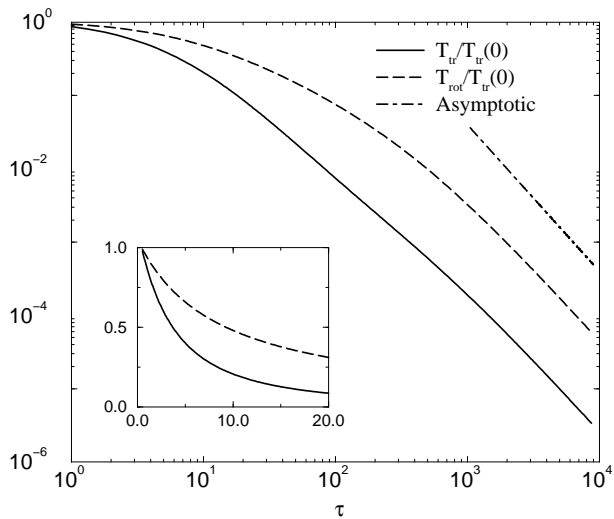, width=8cm}
\caption{Time decay of the translational and rotational
  energy. Parameters were chosen according to:  $k=0.4$,
  $\epsilon=0.4$, $\beta=-0.6$ and $T_{\rm tr}(0)=T_{\rm rot}(0)=20$. 
We have introduced a dimensionless time argument $\tau= t \gamma
  \sqrt{T_{\rm tr}(0)}$. 
The dashed--dotted line indicates the asymptotics}
\label{fig1}
\end{figure}

As we have seen above the energies decrease linearly for short times.
For large times both,
translational and rotational energy fall off algebraically 
like $t^{-2}$ with however different
amplitudes. The amplitudes can be determined analytically, e.g. by solving
the differential equation for
\begin{equation}
\frac{dT_{tr}}{dT_{rot}}=
\frac{-T_{tr}\left((1-\epsilon^2)/4+\eta_t(1-\eta_t)\right)
+T_{rot}\eta_t^2/k}{T_{tr}\eta_t^2/k-T_{rot}(1-\eta_t/k)\eta_t/k}
\end{equation}
This equation is solved by a constant ratio
$T_{tr}/T_{rot}=-c_1/c_2+\sqrt{1+(c_1/c_2)^2}$ 
with $c_1,c_2$ given in terms of $\epsilon,\beta$ and $k$ 
by $c_1=(1-\epsilon^2)/4+(1-\beta^2)(k-1)/(4k+4)$ and 
$c_2=2k(1+\beta)^2/(2+2k)^2$ in agreement with ref.\cite{Goldshtein}.
The crossover time $t_0$ 
between the linear regime and the algebraic decay is determined by 
$\gamma$ and $T_{\rm tr}(0)$ according to 
$t_0^{-1} \propto \gamma \sqrt{T_{\rm tr}(0)}$.
The full solution\cite{Kamke} has been obtained by numerical
integration of eqs.(\ref{coupletr},\ref{couplerot}) and is shown in
Fig. (\ref{fig1}) for
$k=0.4$, $\gamma=1$, $\epsilon=0.4$, $\beta=-0.6$ and
$T_{\rm tr}(0)=T_{\rm rot}(0)=20$.
The asymptotics
is indicated by a dashed--dotted line and the inset is a
blow up of the short time behaviour. 

\begin{figure}[hbt]
\center  \epsfig{file=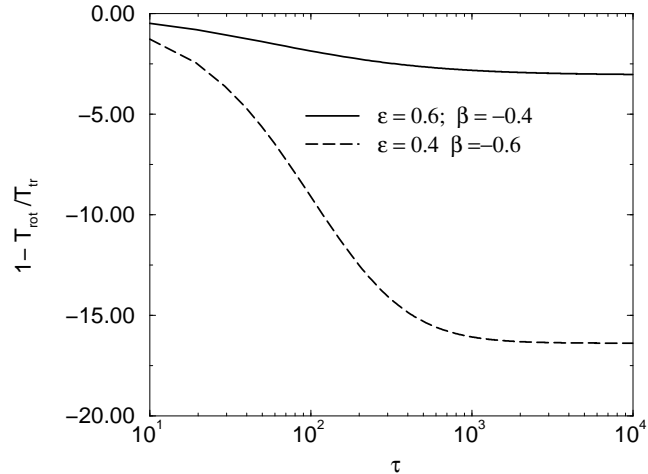, width=8cm}
\caption{Ratio $1-T_{\rm rot}/T_{\rm tr}$ as a function of time
with $k=0.4$, $\gamma=1$, $T_{\rm tr}(0)=T_{\rm rot}(0)=100$,
  $\epsilon=0.6$, $\beta=-0.4$ and $\epsilon=0.4$, $\beta=-0.6$. Unit
  of time was chosen as in Fig. (\ref{fig1}).}
\label{fig2}
\end{figure}

The deviation of the ratio of
rotational and translational energy from $1$, i.e.
$1-T_{\rm rot}/T_{\rm tr}$, is shown in Fig. (\ref{fig2}) for two different values of
$\beta$ and $\epsilon$.

Cooling of a granular gas has been investigated by several groups 
\cite{McNamara2,Goldhirsch} modelling collisions with a normal
coefficient of restitution. 
The $t^{-2}$ behaviour of the total energy has been confirmed for a wide
range of parameters such that the system is stable to shear
fluctuations\cite{Goldhirsch}  and inelastic
collapse.
Luding\cite{Luding} has considered a more detailed model of inelastic
collisions 
including tangential restitution and Coulomb's law of friction. 
It is straightforward
to generalise the model to include Coulomb friction, one just has to
modify the updating rules (eq. (\ref{freddy})) accordingly. Driving the
system with a vibrating wall is an open problem so far.  One may also
generalise the approach to nonspherical objects, e.g. hard rods or
needles\cite{Huthmann}, for which elastic collisions have been
analysed by Frenkel and Maguire\cite{Frenkel}.

We wish to thank Timo Aspelmeier and Kurt Broderix for 
useful discussions and one of us (M.H.)
the Land Niedersachsen for granting a scholarship.

Note added in proof: After this paper was completed we learnt about
related work by  S. McNamara and S. Luding 
who also study the ratio of translational and rotational energy
in the HCS.


\begin{references}
\bibitem{fuss}
 In some calculations it is helpful to know the evolution backwards
 in time. In that case  the momenta after collision are given
 and one wants to know the momenta before collision. 
 These are obtained by replacing
 the parameters in eq. (\ref{freddy}) according to
 $  \beta  \rightarrow \frac{1}{\beta}$ and $ \epsilon  \rightarrow 
 \frac{1}{\epsilon}$.
 With this replacement the momenta before collision are given by the 
 right hand side of eq.(\ref{freddy}) 
 in terms of the momenta after collision.
 These collision rules can then be used to construct an operator 
${\cal L}_{-}$, which describes the evolution backwards in time, but
in general ${\cal L}_{-}$ is  not the adjoint of ${\cal L}_{+}$ (see
 also \cite{Noije}).
\end{references}
\end{document}